\documentclass[
showpacs,floatfix,
aps,prl,amsmath,
twocolumn,
superscriptaddress,
]{revtex4}

\usepackage[varg]{txfonts}

\usepackage{graphicx}
\usepackage{bm}

\begin{document}


\title{Aharonov-Bohm effect as a probe
of interaction between magnetic impurities}

\author{Victor M. Galitski} \affiliation{Condensed Matter Theory Center,
  Department of Physics, University of Maryland, College Park, MD
  20742}
\author{Maxim G. Vavilov} \affiliation{
   Center for Materials Sciences and Engineering,
  Massachusetts Institute of Technology, Cambridge, MA 02139}
\author{Leonid I. Glazman} \affiliation{Theoretical Physics Institute,
  School of Physics and Astronomy, University of Minnesota,
  Minneapolis, MN 55455}

\begin{abstract}
  We study the effects of the RKKY interaction between magnetic impurities
  on the mesoscopic conductance fluctuations of a metal ring with
  dilute magnetic impurities. At sufficiently low temperatures and
  strong magnetic fields, the loss of electron coherence occurs mainly
  due to the scattering off rare pairs of strongly coupled magnetic
  impurities. We establish a relation between the dephasing
  rate and the distribution function of the exchange
  interaction within such pairs. In the case of the RKKY exchange interaction,
  this rate exhibits  $ 1/B^2$ behavior in strong magnetic fields. We
  demonstrate that the Aharonov-Bohm conductance oscillations may be used as
  a probe of the distribution function of the exchange interaction
  between magnetic impurities in metals.

\end{abstract}


\maketitle

The exchange interaction between magnetic impurities
and itinerant electrons drastically affects electron
transport in metals.
Recent experiments \cite{Pierre1,Pierre2,Mohanty,Bauerle}
have revealed a significant effect of
magnetic impurities on the coherent transport
(weak localization and conductance fluctuations)
and electron energy relaxation in metals with even a tiny
concentration of magnetic impurities.
In measurements \cite{Pierre1},  the amplitude of
the ``$h/e$'' Aharonov-Bohm (AB) oscillations in mesoscopic metal rings showed
a strong dependence on the value of the applied field $B$,
with the oscillations being suppressed at low fields and restored at
relatively high fields. This observation is consistent
with the picture of uncorrelated magnetic impurities frozen by
a high magnetic field~\cite{Falko,VG}.

The exchange coupling between magnetic impurities and electrons also gives rise
to the Ruderman-Kittel-Kasuya-Yosida (RKKY) interaction between spins
of magnetic impurities~\cite{Kittel}.  At sufficiently low
temperatures the RKKY interaction between magnetic impurities affects
the electron kinetics in metals~\cite{VGL}. The observation of the
effect of interaction between magnetic impurities was pursued in
experiments \cite{Bauerle}, where the temperature dependence of the
resistivity and the electron phase relaxation rate in diluted magnetic
alloys were investigated.

In this paper, we study the effect of the RKKY interaction on the
AB oscillations of the conductance~\cite{ABeffect,AS}
of mesoscopic metal rings containing magnetic impurities.
We show that the amplitude
of the AB
oscillations may be used as an experimental tool
to probe the statistics of the exchange interaction between magnetic
impurities.  Pairs of magnetic impurities coupled by the exchange
interaction $V$ comparable to the Zeeman energy $g\mu B$ of an
impurity in a magnetic field $B$, play a special role (here $g$ is the
$g-$factor of a magnetic impurity and $\mu$ is the electron Bohr's
magneton).  The dynamics of such resonant pairs is not quenched by
the field $B$, while the other spins are frozen. The resonant pairs
have the strongest effect on the amplitude of the AB conductance
oscillations. The oscillations are characterized by the following correlation
function~\cite{AS},
\begin{equation}
\left\langle \, G_{\Phi}(B) G_{\Phi+\Delta\Phi}(B) \, \right\rangle
=\frac{\alpha e^4}{\pi^2\hbar^2}\sum_{k=0}^{\infty}
{\cal A}_k(B)\cos\left(2\pi k\frac{\Delta\Phi}{\Phi_0}\right),
\label{eq1}
\end{equation}
where $G_{\Phi}(B)$ is the conductance of the ring in the presence of
a magnetic flux $\Phi$, $\Phi_0 = hc/e$ is the flux quantum,
and $\alpha$ is a geometry-dependent numerical factor.
The strength of the AB effect is determined by the amplitudes ${\cal
  A}_k(B)$ with $k\neq 0$. In the limit of low temperatures and high
  magnetic fields, we find
\begin{equation}
{\cal A}_k(B)\propto
\exp\left(-\frac{kL}{L_\varphi}\right),\quad
\frac{1}{L^2_\varphi}\propto T P(g\mu B).
\label{eq2}
\end{equation}
Here $L$ is the ring circumference, $T$ is the
electron temperature  and $P(g\mu B)$ is the density of magnetic
impurity pairs coupled by exchange
interaction of strength $V=g\mu B$. Thus, the measurement of the AB
amplitudes ${\cal A}_k(B)$ allows one to ``scan'' the distribution
function $P(V)$ of the interaction strength between magnetic
impurities. The applicability of Eq.~(\ref{eq2}) requires
$T\ll\overline{V}\ll g\mu B$, where $\overline{V}$ is the typical strength of
the interaction between magnetic impurities.

The AB conductance oscillations,
being an interference phenomenon~\cite{ABeffect,AS},  are very sensitive to any
changes in disorder realization \cite{obzor,Levy}. Particularly, temporal
changes in the disorder configuration during the measurement process
suppress the amplitudes ${\cal A}_k$ of the conductance correlation
function~(\ref{eq1}).  This suppression is a result of the conductance
self-averaging during the measurement. A possible mechanism for such
suppression comes from the evolution of spin configuration in the system
of magnetic impurities~\cite{Falko} on the time scale defined by
impurity spin relaxation. This scale is much shorter than the
conductance measurement time. However, if the spins of magnetic
impurities are quenched, they do not suppress the amplitude of the
conductance fluctuations.

A way to quench the spin dynamics of magnetic impurities is to apply a
magnetic field. In the system of non-interacting magnetic impurities,
the magnetic field $B$ yields an exponential suppression of the
spin-flip scattering rate $\gamma_{\rm s}$: if the Zeeman energy
splitting exceeds temperature, $g \mu B \gg T$, then $\gamma_{\rm s}
\propto \exp{\left[ - {g \mu B/T} \right]}$, see~\cite{Falko,VG}.
Interaction between magnetic impurities may weaken the spin quenching
by magnetic field, if the interaction strength between impurities is
comparable with the Zeeman splitting. We consider strong magnetic
fields when the majority of spins are frozen by the applied field. Due
to random positions of magnetic impurities, there exist strongly
interacting pairs for which the dynamics is not quenched. Such pairs
effectively act as two-level systems, and provide the main contribution
to the suppression of the AB conductance oscillations.  We note that in
strong magnetic fields (the Zeeman splitting exceeds the typical
strength of the interaction between magnetic impurities) the
probability to find three or more strongly coupled magnetic impurities
is much smaller than the probability of finding a strongly coupled
pair.  In evaluating $\gamma_{\rm s}$, we account only pairs of magnetic
impurities, following Refs.~\cite{LMK,LK}.

The competition between the antiferromagnetic exchange $V>0$ and the Zeeman
splitting results in the degeneracy of spin states of impurity pairs
for which the condition $V=g\mu B$ is satisfied. The spin dynamics of
such pairs is not quenched, and they contribute to the
electron dephasing. The remaining impurity spins are frozen by the
applied magnetic field or interactions and do not suppress the
conductance fluctuations (we note that the dynamics of
ferromagnetically coupled spins, $V<0$, is quenched even stronger than
the dynamics of non-interacting spins). Therefore, the amplitudes
${\cal A}_k(B)$, see Eq.~(\ref{eq1}), are determined by the magnetic
impurity pairs characterized by the exchange interaction strength $V$
of the order of the Zeeman splitting: $\left| V - g\mu B \right|
\lesssim T$.

We focus on the
limit of strong spin-orbit scattering (the spin-orbit scattering rate
is supposed to be much larger than the scattering rate off magnetic
impurities). The interaction of an electron with a magnetic impurity is
described by the exchange Hamiltonian ${\cal H}_{\rm ex} = {\cal J}
\hat{\bm {S}} \hat {\bm{\sigma}}$, where ${\cal J}$ is the exchange
coupling constant, $\hat{\bm {S}}$ is the spin of a magnetic impurity,
and $\hat{\bm{\sigma}}$ is the electron spin density. The exchange
coupling leads to the effective RKKY interaction between magnetic
impurities.
If the distance between the impurities in a pair is
smaller than the spin-orbit scattering length $L_{\rm so}$, the
coupling between the impurity spins $\hat {\bm S}_{1,2}$
is isotropic~\cite{Spivak} and the corresponding Hamiltonian is
\begin{equation}
\label{H_2}
{\cal H}_{\rm S}^{(2)} = V { \hat{\bm S}}_1 {\hat{\bm S}}_2 +
g \mu B \left[  \hat{S}_{1z} + \hat{S}_{2z} \right].
\end{equation}
Here  $V$ is the RKKY coupling.
The eigenstates of this Hamiltonian $|\xi\rangle = | J, M \rangle$ are
classified by the total spin $J$ and the projection $M$ of the total
spin on the direction of the magnetic field.
The energy spectrum of~Eq.(\ref{H_2}) has the
form
\begin{equation}
E_{J,M} = \frac{V}{2} \left[ J \left( J + 1 \right) - S \left(
S + 1 \right) \right] + g \mu B M.
\label{spektr}
\end{equation}
Below we consider spin-$1/2$ magnetic impurities, in which case there
are four energy levels corresponding to three $J=1$ states and one
$J=0$ state.

We calculate the spin-flip scattering rate off pairs of magnetic impurities
\begin{equation}
\gamma_{\rm s}(\varepsilon)=
\frac{4}{3\tau_{\rm s}}\int dV
P(V)\left[
K(\varepsilon,V)-\langle S_z \rangle^2
\right],
\label{gammas}
\end{equation}
where
$\tau_{\rm s}^{-1} = \left( {3 \pi / 2} \right) n_s \nu {\cal J}^2$ is
the scattering rate off an isolated impurity at zero magnetic field,
$\nu $ is the electron density of states at the
Fermi level and $n_{\rm s}$ is the concentration of magnetic
impurities.  The term $4K (\varepsilon,V)/3\tau_{\rm s}$ in
Eq.~(\ref{gammas}) represents the scattering rate of an electron with
energy $\varepsilon$ off a magnetic impurity, which belongs to a pair
characterized by the exchange interaction strength $V$. This rate can
be obtained from the Fermi golden rule:
\begin{equation}
K(\varepsilon,V)=
\sum_{\xi\xi'}\frac{ e^{-E_\xi/T}}{Z}
\left| \langle\xi|\hat{\bm S}|\xi'\rangle \right|^2
\frac{1+e^{\varepsilon/T}}{1+e^{(\varepsilon-E_\xi+E_{\xi'})/T}},
\label{ImT}
\end{equation}
where $Z=\sum_\xi e^{-E_{\xi}/T}$ is the partition function.  The
polarization $\langle S_z\rangle =\sum_\xi e^{-E_\xi/T}\langle \xi|
\hat{S}_z|\xi\rangle/Z $ of magnetic impurities reduces fluctuations of
impurity spins and also reduces $\gamma_{\rm s}$; this effect is described
by the $4\langle S_z\rangle ^2/3\tau_{\rm s} $ term in
Eq.~(\ref{ImT}). The rate $\gamma_{\rm s}(\varepsilon)$ represents the
result of averaging over the strength of the exchange coupling $V$
with the weight $P(V)$. Here $P(V)$ denotes the distribution function
of the exchange coupling $V$ for pairs of magnetic impurities, which
we discuss later.

We consider Zeeman splitting $g \mu B$ exceeding both
temperature $T$ and the typical inter-impurity coupling
$\overline{V}$.  In this case, the strength of interaction between
magnetic impurities becomes comparable with $g \mu B$ at distances
much shorter than the typical distance between magnetic impurities
$n_{\rm s}^{-1/3}$, which justifies the use of the virial expansion
method. If the characteristic scale for variation of $P(V)$ is larger
than temperature $T$, we obtain from (\ref{gammas}) and (\ref{ImT})
\begin{equation}
\gamma_{\rm s}(\varepsilon)=\frac{T P( g\mu B )}{3\tau_{\rm s}}
\left(
1+ \frac{2 \varepsilon}{T}\coth\frac{\varepsilon}{2T}
\right), \quad |\varepsilon|\ll g\mu B.
\label{gammasVirial}
\end{equation}
The spin flip rate (\ref{gammasVirial}) takes into account the
scattering processes which change the state of impurity spins in
a pair from $J=1$ and $M=-1$ to $J=0$ and {\em vice versa}. Therefore, a
pair of impurity spins coupled with the strength $V-g\mu B \lesssim T$
acts as a two level system~\cite{Galperin}.

We emphasize that, in general, the scattering rate off magnetic
impurities is not a single universal parameter, which determines all
transport properties of an electron liquid with embedded magnetic
impurities. Various properties of the conductance, such as the weak
localization correction~\cite{HLN} and the conductance fluctuations,
are determined by different parameters characterizing scattering off
magnetic impurities (see Ref.~[\onlinecite{VG}] for a detailed
analysis). Here Eq.~(\ref{gammasVirial}) defines the dephasing
rate $\gamma_{\rm s}(\varepsilon)$, which limits the
 amplitude
of the Aharonov--Bohm oscillations, Eq.~(\ref{eq1}).

The amplitudes ${\cal A}_k$ in Eq.~(\ref{eq1})
in the limit of weak scattering off magnetic impurities,
$\gamma_{\rm s}(T) \ll T,\; D/L_{\rm so}^2$,
can be written in the form~\cite{VG}
\begin{equation}
{\cal A}_k=\frac{D^{3/2}}{T^2R^3}\int
\frac{
e^{- k L\sqrt{\gamma_{\rm s}(\varepsilon)/D}}}
{\sqrt{\gamma_{\rm s}(\varepsilon)}}
\frac{d\varepsilon}{\cosh^4(\varepsilon/2T)},
\label{Rk}
\end{equation}
where $D$ is the diffusion coefficient.
Substituting Eq.~(\ref{gammasVirial}) into Eq.~(\ref{Rk}), we find the
amplitudes of the conductance correlation function harmonics. If the
circumference of the ring $L$ is much larger than the coherence length
$L_\varphi $,
then the saddle point approximation is applicable and we find
\begin{subequations}
\begin{eqnarray}
{\cal A}_{k}  & = & \sqrt{\frac{15}{k}}
\frac{L_T^2L_\varphi^{3/2}}{[L/2\pi]^{7/2}}
\exp\left(- k \frac{L}{L_\varphi}\right),
\label{Ak}
\\
\frac{1}{L_\varphi^2} &= &\frac{\gamma_{\rm s}}{D} =
\frac{5T}{3\tau_{\rm s}D}P(g\mu B),
\label{Lphi}
\end{eqnarray}
cf. Eq.~(\ref{eq2}).
Here $L_T=\sqrt{D/T}$ is the thermal length,
$\gamma_{\rm s}=\gamma_{\rm s}(\varepsilon=0)$, and
the coherence length  $L_\varphi$ in a magnetic
field $B$ is determined by the probability $P(V)$ of finding a pair
of two magnetic impurities antiferromegnetically
interacting with each other with the
strength $V=|g\mu B|$, see Eq.~(\ref{gammasVirial}). Therefore, the
amplitude of the conductance fluctuations can be used to probe the
distribution function $P(V)$.  Below we discuss the appropriate
distribution functions for the RKKY interaction in the limits of
strong and weak disorder.
\end{subequations}

The strength of the RKKY coupling is
determined by the electron wave functions at the positions of magnetic
impurities. In a disordered metal, the values of the wave functions 
at distances $r$ larger than the mean free path $l$ are nearly
uncorrelated, and the average value of RKKY coupling over disorder is
exponentially small: $\left\langle V(r) \right\rangle \sim \exp{\left(
    -r/l \right)}$. However, the values of $V(r)$ in a given
configuration of disorder remain $\propto 1/r^3$ even in the ``strong
disorder'' limit $r\gg l$, as one can
see~\cite{Spivak,AbrahamsPRB37,LernerPRB48} from the variance of RKKY
coupling,
$\left\langle V^2(r)\right\rangle=(3/4)C^2 /r^{6}$ with
$C =\nu{\cal J}^2/2\pi$.

We emphasize that in a disordered metal the interaction strength between
two magnetic impurities is not uniquely determined by the distance between these
impurities, but also depends on disorder realization. The full
distribution function $p(V,r)$ of $V(r)$ is not known
even for a metal with $k_Fl\gg 1$. However, from the $1/r^6$
behavior of the variance on the distance between magnetic impurities, we
assume that the distribution function $p(V,r)$ has the following scaling form
$p(V,r)=r^3\zeta(r^3V/C)/C$.
For magnetic impurities randomly positioned in space according to the
Poisson distribution, the probability of finding the nearest neighbor
at a distance $r\ll n_{\rm s}^{-1/3}$ is equal to $ n_{\rm s} $. To
derive the probability distribution function of the RKKY interaction
strength, we average over the distance between magnetic
impurities $r$:
\begin{equation}
\label{Jexa}
P_{\rm dis} (V)
=  \int 4\pi r^2 \frac{r^3 \zeta(r^3 V/C)}{C} dr
 =  \frac{\overline{V} }{V^2},\ \
\overline{V} = \frac{4}{3}\eta \pi C n_{\rm s}.
\end{equation}
Here we introduced the typical interaction strength
$\overline{V}$ between magnetic impurities at the distance $n_{\rm s}^{-1/3}$
and $ \eta = \int_0^{\infty} x \zeta(x) dx$ is a numerical factor.
We emphasize that Eq.~(\ref{Jexa}) is applicable for
impurities at $r\ll n_{\rm s}^{-1/3}$, {\emph i.~e.},
for $V\gg \overline{V}$, where the virial
approximation is justified.
According to Eq.~(\ref{gammasVirial}), the corresponding scattering rate is
\begin{equation}
\label{sf_result}
\gamma_{\rm s} =  \frac{5 }{3  \tau_{\rm s}}
\frac{T \overline{V}}{\left( g \mu B \right)^2},
\end{equation}
in sharp contrast to the exponential decay rate for isolated magnetic
impurities in a strong magnetic field $B$~\cite{Falko,VG}.

We note here that the scattering rate off non-interacting magnetic
impurities calculated to fourth order in the exchange coupling
${\cal J}$ also has a power law dependence on the applied magnetic
field~\cite{VG}: $\gamma_{\rm s}^{(4)} \propto T^2 / (g \mu B)^2$.  However,
this rate becomes smaller than the
rate $\gamma_{\rm s}$, Eq.~(\ref{sf_result}),
at $T\lesssim \overline{V}$.  Therefore,  in the low-temperature
and strong magnetic field limit, $T \ll   {\overline{V}} \ll g \mu
B$,
the amplitudes ${\cal A}_k(B)$ of the ``$h/e$'' Aharonov--Bohm
oscillations, Eq.~(\ref{eq2}), are
determined by electron scattering off magnetic impurities coupled
in pairs. The corresponding coherence length $L_{\varphi}$ and dephasing rate
$\gamma_{\rm s}$ are
given by Eqs.~(\ref{Lphi}) and (\ref{sf_result}), respectively.

As the magnetic field increases, suppression of the AB amplitudes is
determined by impurity pairs with the typical separation
of order $r(B) = \left[ C  / \left( g \mu B
\right)\right]^{1/3}$, where
the factor $C  =\nu {\cal J}^2/2\pi$ is determined by the renormalized
exchange constant ${\cal J}$ due to the Kondo effect and can be
estimated as
\begin{equation}
\label{C}
C(r) = {2 \over \pi \nu} \ln^{-2} \left[ \frac{{\rm max}\,
\left\{ v_{\rm F}/r, g \mu B, T \right\}}{ T_{\rm K}}  \right].
\end{equation}
A question arises: what happens when the distance becomes smaller
than the elastic mean free path $l$?
At such distances, the
mesoscopic fluctuations of the
RKKY exchange interaction are negligible, therefore
\begin{equation}
V(r)=\frac{C}{r^3}\cos 2p_{\rm F} r.
\end{equation}
The corresponding distribution function $P(V)=P_{\rm c }(V)$,
\begin{equation}
P_{\rm c } (V) =  \int\limits d^3{\bf r}\,
\delta\left[ V - \frac{C}{r^3} \cos \left( 2 p_{\rm F} r \right) \right].
\end{equation}
is non-monotonic and
possesses cusps $1/\sqrt{V_n-V}$ at the
maxima of the RKKY interaction $V_n = \left( 8 / \pi^2\right) \left(
E_{\rm F} / n^3 \right) \ln^{-2} \left[ E_{\rm F} / n T_{\rm K}
\right]$; here
the maxima are labeled with the index
$n \sim \left( E_{\rm F} / g \mu B \right)^{1/3} \gg 1$ and
$E_{\rm F}$ is the Fermi energy.
These singularities should
reveal themselves in high-field measurements [as defined below in
Eq.~(\ref{cond})] of the conductance correlation function, being
however partially smeared out by temperature $T$ and non-magnetic
disorder.

\begin{figure}[htbp]
\centering
\includegraphics[width=3.05in]{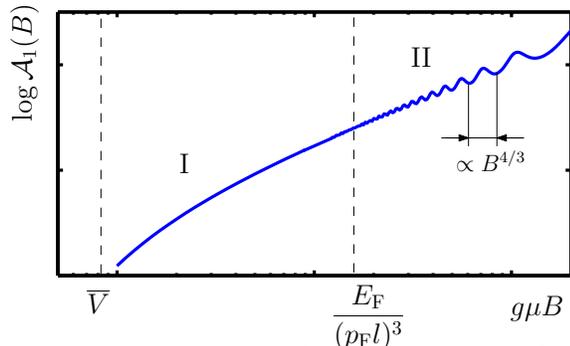}
\vspace*{-0.2in}
\caption{ Schematic log-log plot of the amplitude ${\cal A}_1(B)$
of the AB oscillations as a function of the Zeeman splitting $g\mu B$.
Two regimes are shown: I) interacting pairs of magnetic
impurities separated by distances larger than the mean free path $l$
described by Eqs.~(\ref{Ak}),  (\ref{Lphi}), and (\ref{sf_result}); 
II) interacting pairs of magnetic impurities separated by distance 
smaller than the mean free path $l$. }
\label{fig1}
\end{figure}

The oscillations of the amplitude of the conductance correlation
function due to the singularities of $P(V)$ can be resolved if the
distance between the neighboring maxima $\Delta V_n=V_{n}-V_{n+1}$ is
larger than the temperature; this condition
is satisfied at $3V_n/n\gg T$.  The
singularities in $P(V)$ are not smeared by disorder if
the magnetic impurities in a pair are
separated by distances smaller than the mean free
path. The latter condition yields $n\ll lp_{\rm F}/\pi$.
Keeping in mind that the
distance between magnetic impurities being probed is to be much larger
than the lattice spacing, we obtain the following set of conditions
for the observability of
the modulation of the amplitudes ${\cal A}_k(B)$ of the AB conductance
oscillations:
\begin{equation}
\label{cond}
 {\rm max}\,\left\{\left({\frac{T}{E_{\rm F}}}\right)^{3/4}\!\!\! ,
\frac{1}{(p_{\rm F}l)^3} \right\} \ll
\frac{g \mu B}{E_{\rm F}} \ll   \ln^{-2} \left[ {E_{\rm F} \over T_{\rm K}}
\right].
\end{equation}
 Estimating $E_{\rm F}
\sim 10^4\,{\rm K}$ and $p_{\rm F} l \sim 100$, we conclude that the
magnetic field can be as low as $g \mu B \sim 1{\rm K}$ and the
temperature as high as $T \sim 10{\rm mK}$, {\em i.~e.}, the effect is
within the experimentally accessible range (see {\em e.~g.},
Ref.~[\onlinecite{Bauerle}]).  To identify the
oscillations, one should study the dependence of
the period and amplitude of oscillations in $\gamma_{\rm s}$ on the applied
magnetic field $B$. The period increases as $B^{4/3}$ and
the amplitude grows as
$\delta\gamma^{\rm osc}_{\rm s} / \gamma_{\rm s} \propto  B^{1/3}$
with increasing magnetic field $B$.

In Fig. 1, we schematically present the dependence of amplitudes $A_k$
of the AB oscillations on the energy of Zeeman splitting $g\mu B$
for magnetic impurities, assuming $g\mu B\gg\bar V$.
If the distance between two magnetic impurities interacting with the
strength $g\mu B$ is larger than the mean free path $l$, the amplitude
of the AB conductance oscillations ${\cal A}_k(B)$ is a monotonic function of the
applied magnetic field, described by
Eqs.~(\ref{Ak}) and (\ref{sf_result}), see region I in Fig.~\ref{fig1}. At stronger magnetic
fields, defined by Eq.~(\ref{cond}), the amplitudes ${\cal
A}_k(B)$ become non-monotonic, see region II in Fig.~\ref{fig1}.

In summary, we have shown that
the amplitude of mesoscopic conductance fluctuations in
a dilute magnetic alloy in a high magnetic field is determined by the
density of strongly coupled pairs of magnetic impurities.  Measurements of the
Aharonov--Bohm oscillations in mesoscopic rings may be used as a probe
of the distribution function of the interaction strength between
magnetic impurities. High-field measurements of the conductance
correlation function can provide a direct insight into the oscillatory
nature of the RKKY interaction.

The authors  are grateful to L.~L{\'e}vy for turning their attention to the problem 
and  C.~B\"auerle and I.~Lerner for discussions.
V.~G. was supported by the US-ONR, LPS, and DARPA; M.~V. was supported
in part by the MRSEC Program of the National Science Foundation under
award DMR 02-13282; L.~G. was supported by NSF grants
DMR02-37296 and EIA02-10736 (University of Minnesota).
\vspace*{-0.2in}

\bibliography{RKKY2}

\end{document}